\documentclass{amsart}
\usepackage{graphicx} 
\usepackage{natbib}
\usepackage{amssymb}
\usepackage{proof}
\usepackage{mathrsfs}

\newcommand{\sat}{\Vdash}

\newcommand*{\rn}[1]  {\ensuremath{\mathsf{#1}}}
\newcommand*{\irn}[2][]  {#2_\rn{#1{I}}}
\newcommand*{\ern}[2][]  {#2_\rn{#1{E}}}

\newcommand{\base}[1]{\mathscr{#1}}
\newcommand{\supp}[1]{\sat_{#1}}

\title{A Note on the Practice of Logical Inferentialism: The State-Effect Interpretation, Definitional Reflection, 
and Completeness}

\author{Alexander V. Gheorghiu}

 \address{Department of Computer Science\\
 University College London\\
 London WC1E 6BT,~UK}
 \email{alexander.gheorghiu.19@ucl.ac.uk}

\author{Tao Gu}

 \address{Department of Computer Science\\
 University College London\\
 London WC1E 6BT,~UK}
 \email{tao.gu.22@ucl.ac.uk}

\author{David J. Pym}
 \address{Institute of Philosophy,
 University of London, 
 London WC1E 7HU, UK \and \newline Department of Computer Science and Department of Philosophy,
 University College London,
 London WC1E 6BT, UK}
 \email{d.pym@ucl.ac.uk}

\begin{document}

\maketitle

The prevailing paradigm of semantics in logic is \emph{denotationalism}, wherein the meaning of logical structures is grounded in reference to abstract algebraic structures, exemplified by \emph{model-theoretic semantics} (M-tS). An alternative paradigm is \emph{inferentialism}~\citep{brandom2009articulating}, wherein meaning emerges from rules of inference. It may be viewed in the context of the \emph{meaning-as-use} principle \citep{Wittgenstein}, with `use' understood in logic as `inference'.

For instance, consider the statement `Luna is a vixen', which intuitively means `Luna is a fox' \emph{and} `Luna is female'. In inferentialism, its meaning is given by the following rules:
\[ 
\begin{array}{c}
\infer{\text{Luna is a vixen}}{\text{Luna is a fox} & \text{Luna is female}} \qquad
\infer{\text{Luna is female}}{\text{Luna is a vixen}} \qquad \infer{\text{Luna is a fox}}{\text{Luna is a vixen}} \\
\end{array}
\]
These merit comparison with the laws governing conjunction ($\land$) \citep{Gentzen}, which justify the `and' above,
\[
\infer[\irn \land]{\phi \land \psi}{\phi & \psi} \qquad
\infer[\ern{{\land_1}}]{\phi}{\phi \land \psi} \qquad \infer[\ern{{\land_2}}]{\psi}{\phi \land \psi} 
\]

\emph{Proof-theoretic semantics} (P-tS) provides a mathematical expression of inferentialism \citep{SEP-PtS}. Unlike M-tS, P-tS views \emph{proofs}
 as the basis of meaning (as opposed to \emph{truth}). Below, we propose an analysis of P-tS that explicates the semantic clauses for various standard logics. In particular, we justify the P-tS for \emph{intuitionistic propositional logic} (IPL) by~\cite{Sandqvist2015IL} (cf.~\cite{Piecha2019incompleteness}), and its subsequent developments by  \cite{gheorghiu2023imll} and \cite{gu2023prooftheoretic}.

Heuristically, logical constants are \emph{defined} by their introduction rules; for instance, $\irn \land$ can be understood as \emph{defining} $\land$. A central challenge in P-tS is to formalize this heuristic rigorously. However, merely providing introduction rules for a logic is insufficient for a complete definition; one requires elimination rules. Notably, the choice of elimination rules is not arbitrary \citep{Prior61}, leading to the study of \emph{harmony} \citep{Tennant1978} and \emph{inversion}   \citep{Prawtiz1965} in P-tS.

Notwithstanding these challenges, the elimination rules in the natural deduction systems on which much of P-tS has been based --- for example, \textsc{NJ} \citep{Gentzen} --- are thought of as being \emph{consequences} of their corresponding introduction rules. Indeed, a central approach in P-tS has been to witness and exploit this relationship~\citep{SchroederHeister2006}. One way to characterize the relationship is through the following idea:
\begin{quote}
    The introductions represent the `state' required for a logical constant to be asserted, and the corresponding eliminations represent the `effect' of that constant.  
\end{quote}
This \emph{state-effect} duality is useful for explicating the differences between M-tS and P-tS. Of course, the `effect' of a logical constant must be warranted by the `state' it signifies; whence, harmony and inversion principles. 

Some explanation is required for what we mean by state and effect. They correspond to two perspectives on logic, the \emph{ontological} and the \emph{teleological}.

The ontological perspective views logic as the study of properties of objects \citep{sep-logic-ontology} and `state' refers to the state of objects. This view is the foundation for M-tS and the clauses of the logical constants are, therefore, closely related to their corresponding introduction rules. For example, in the standard possible-world semantics for IPL by \cite{kripke1965semantical}, the clause for disjunction closely mirrors the introduction rule $\irn[1] \lor$ and $\irn[2] \lor$ \citep{Gentzen},
\[
\mathfrak{M}, w \sat \phi \lor \psi \qquad \mbox{iff} \qquad \mbox{$\mathfrak{M}, w \sat \phi $ or $\mathfrak{M}, w \sat  \psi$}
\]
The correctness of the elimination rules
follows from `backward' reasoning on states, which are represented by possible worlds \citep{gheorghiu2023semantical,gheorghiu2023acs,gheorghiu2024thesis}.

In contrast, we shall call the `teleological' perspective on logic the view that is concerned with understanding \emph{reasoning}. \cite{tennant1978entailment} exemplifies this view on logic in the `consequentialist' reading of entailment. By `effect' we mean what reasoning is rendered possible by the assertion of the logical constants. This explains why clauses of the semantics typically mirror the elimination rules.  For example, in the work on IPL by  \cite{Sandqvist2015IL}, the clause for disjunction ($\lor$) closely mirrors the elimination rule $\ern \lor$ \citep{Gentzen} restricted to atoms,
\[
\begin{array}{lcl}
\supp{\base{B}} \phi \lor \psi & \mbox{iff} & \mbox{for any $\base{C} \supseteq \base{B}$ and any atom $p$,}\\ & & \mbox{if $\phi\supp{\base{C}} p$ and $\psi\supp{\base{C}} p$, then $\supp{\base{C}} p$}
\end{array}
\]
The details --- such as the significance of $\base{B}$ and $\base{C}$ --- are not essential for this discussion. The point is that forcing a clause informed by a state-based view yields incompleteness; that is, \cite{Piecha2015failure} have shown that using
\[
\supp{\base{B}} \phi \lor \psi \qquad \mbox{iff}  \qquad \supp{\base{B}} \phi \qquad \mbox{or} \qquad \supp{\base{B}} \psi
\]
does not give a semantics for IPL.  The correctness of the introduction rules follows from the transitivity of the judgment relation.

In the practice of logical inferentialism, while we treat introduction rules as definitions, our central concern is to understand the effects implied by a state. More broadly, how one transitions from an ontological interpretation to a teleological one. We argue that \emph{Definitional Reflection} (DR) \citep{Hallnas1987,Hallnas1990,Hallnas1991,SchroederHeister1993} plays an essential role:
\begin{quote}
``Whatever follows from all the defining conditions of an assertion follows from the assertion itself."
\end{quote}
Not only does DR offer a mathematically elegant solution to these issues but also embodies the intended conceptual shift.

Indeed, DR gives one approach for generating elimination rules from introduction rules. For example, let $\irn[1] \lor$ and $\irn[2] \lor$ be the `definition' of disjunction ($\lor$),
\[
\infer[\irn{{\lor_1}}]{\phi \lor \psi}{\phi} \qquad \infer[\irn{{\lor_2}}]{\phi \lor \psi}{\psi}
\]
Hence, the defining conditions of $\phi \lor \psi$ are $\phi$ and $\psi$.  Whence, by DR, whatever follows from both $\phi$ and $\psi$ also follows from $\phi \lor \psi$. This motivates $\ern \lor$,
\[
\infer[\ern{\lor}]{\chi}{
\phi \lor \psi &
\deduce{\chi}{[\phi]} & \deduce{\chi}{[\psi]}
}
\]
as required. Observe that $\ern[1]{\land}$ and $\ern[2]{\land}$ are together obtained from and equally expressive as the generalized elimination rule for $\irn \land$ \citep{schroeder1984natural}.This is all very well as a mechanism, but why ought we accept it? 

According to \cite{SchroederHeister1993}, DR follows from reading the introduction rules as expressing \emph{all} conditions for the assertion to be made. Essentially, this embodies a form of \emph{closed-world assumption} \citep{Reiter1981}, which in the form \emph{negation-as-failure} \citep{Clark1978} has been studied more directly in the context of the P-tS of IPL \citep{Alex:NAF}. In this sense, it is the completion \citep{Clark1978} of the inference system afforded by the introduction rules; in the state-effect reading proposed above, the generalized elimination rules represent the \emph{maximal} effect of a logical constant corresponding to its introduction rules.

This justifies DR as a general phenomenon. It remains to show that it represents the intended conceptional shift from the ontological to the teleological perspective when doing logical inferentialism. 

\cite{goldfarb2016dummett} has given an alternative  sound and complete inferential semantics of IPL that is based on the ontological view of logic where the \emph{objects} are `canonical' proofs in the Dummett-Prawitz sense \citep{SchroederHeister2006}. In this case, canonical proofs act as states, and the introduction rules act as \emph{constructors}, yielding clauses based on introduction rules. \cite{Alex:PtV-BeS} have shown that moving from constructors to constructiveness --- that is, from states to effects --- in the approach by \cite{goldfarb2016dummett} recovers the semantics by \cite{Sandqvist2015IL}. The move is conceptionally simple and best illustrated by an example:
\begin{quote}
Consider the judgment $\phi \lor \psi \vdash \psi \lor \phi$. We must show that on the assumption that there is a valid argument for $\phi \lor \psi$, there is a valid argument for $\psi \lor \phi$. Without loss of generality, a valid argument for $\phi \lor \psi$ ends by a constructor $\irn[1] \lor$ or $\irn[2] \lor$. Hence, a valid argument exists for either $\phi$ or $\psi$. Whence, by the use of a constructor $\irn[1] \lor$ or $\irn[2] \lor$,  there is a valid argument for $\psi \lor \phi$, as required. 
\end{quote}
Observe that we have implicitly used DR: we have argued from \emph{all the defining conditions} of the assertion $\phi \lor \psi$ that we can conclude $\psi \lor \phi$.

To summarize, adopting the state-effect interpretation of introduction and elimination rules aids in delineating the distinctions between P-tS and M-tS. This approach offers a unified understanding of the practice of logical inferentialism. Additionally, DR emerges as a central tenet in transitioning from a state-based semantics to an effect-based semantics of a logical system. It not only works at a mathematical level, but actually arises naturally as one moves between the perspectives. 

\bibliographystyle{agsm}
\bibliography{bib}
\end{document}